 \documentclass[11pt]{article}

 \pdfoutput=1

\usepackage{graphicx} 
\usepackage{amsmath}

\usepackage{enumitem}

\usepackage[francais,english]{babel}

\usepackage{float}
\usepackage{ifthen}
\usepackage{hyperref}

\begin{document}


\title{Moist-entropic vertical adiabatic lapse rates: the standard cases and some lead towards inhomogeneous conditions.}

\author{by Jean-Fran\c{c}ois Geleyn and Pascal Marquet. {\it M\'et\'eo-France.}}


\date{14th of January, 2014}

\maketitle


\begin{center}
{\em A note published in the \underline{WGNE Blue-book} in 2012.} \\
{\em \underline{Corresponding addresses}: jean-francois.geleyn@meteo.fr / pascal.marquet@meteo.fr}
\end{center}
\vspace{1mm}


\begin{abstract}
This note is a companion of Marquet and Geleyn (2013 \url{http://arxiv.org/abs/1401.2379}
{\tt arXiv:1401.2379 [ao-ph]}), where adiabatic lapse rates $\Gamma_{ns}$ and $\Gamma_{sw}$ are derived for non-saturated ($\Gamma_{ns}$) or saturated ($\Gamma_{sw}$) parcel of moist-air.
They are computed in terms of the vertical derivative of the moist-air entropy potential temperature $\theta_s$ defined in Marquet (2011).
The saturated value $\Gamma_{sw}$ is rewritten in this note so that a more compact formulation is obtained.
The new formulation for $\Gamma_{sw}$ is expressed in term of a weighting factor $C$.
This factor may represent the proportion of an air parcel being in saturated conditions.
\end{abstract}


 \section{Introduction.} 
\label{section_INTRO}

In a recent paper, Marquet (2011) proposed a new moist-entropic potential temperature $\theta_s$, linked to the second law of thermodynamics through its full equivalence to the specific moist entropy $s$, {\it i.e.\/} with consideration of the ``dry air'' and ``water species'' subparts of the atmospheric parcel, of specific content $q_d$ plus the total water specific content $q_t=1-q_d=q_v+q_l+q_i$. 
The likely advantage of $\theta_s$ with respect to earlier proposals is that it is both Lagrangian-conservative and tractable in mixing processes. 

Given the obvious links between any kind of potential temperature and vertical adiabatic lapse rates, we elected to do the analytical computation of such lapse rates on the basis of parcels keeping $\theta_s$ constant. 
This can be done without approximation only for the cases of
\begin{itemize}[labelsep=5mm,listparindent=0mm,parsep=0cm,itemsep=1mm,topsep=0cm,rightmargin=2mm]
\item  no condensed phase at all (named here ``\underline{non-saturated}'', rather than using the ambiguous ``\underline{dry}''); and 
\item  fully saturated conditions (named here ``\underline{saturated}'').
\end{itemize}
 
Beware that we shall consider here only saturation with respect to liquid water, the extension to ice water conditions being rather straightforward. 
In fact, the results presented below were obtained with the even more ambitious goal to look at the vertical stability under any (neutral or not) conditions, see Marquet and Geleyn (2013). 
But we shall concentrate here on the adiabatic lapse rates, going for this to further details than in the above-mentioned paper.

Despite the apparent complexity of the analytical formulation for $\theta_s$
\begin{align}
  {\theta}_{s} 
  & =
        \: T \;
        \left( \frac{p_0}{p}\right)^{R_d/c_{pd}}
        \:
        \exp \left( - \:
                    \frac{L_{vap}(T)\:q_l}{{c}_{pd}\:T}
                \right)
      \; \exp \left( \Lambda_r \:q_t \right) \;
  \nonumber \\
  & \quad \times \;
        \left( \frac{T}{T_r}\right)^{\lambda \:q_t}
        \left( \frac{p}{p_r}\right)^{-\kappa \:\delta \:q_t}
    \left(
      \frac{r_r}{r_v}
    \right)^{\gamma\:q_t}
     \;
      \frac{(1+\eta\:r_v)^{\:\kappa \: (1+\:\delta \:q_t)}}
          {(1+\eta\:r_r)^{\:\kappa \:\delta \:q_t}}
  \: , \label{def_THs}
\end{align}
the results derived in Marquet and Geleyn (2013) in terms of adiabatic lapse rates are beautifully compact.

The non-saturated formulation is given by Eq.(11) in Marquet and Geleyn (2013).
It can be written as
\begin{align}
  \Gamma_{ns} &  \:  = \:
     \frac{g}{c_p}
  \: . \label{def_Gamma_ns}
\end{align}
This was an expected result, with $c_p$ obviously depending on the parcel's composition ($q_d$, $q_v$, $q_l$, $q_i$).

The saturated formulation given by Eqs.(16)-(18) in Marquet and Geleyn (2013) can be written as
\begin{align}
  \Gamma_{sw} &  \: = \:
     \frac{g}{c_p}
          \: \left[ \: 
     \frac{ 1 \: + \:
      \left( 1 + \eta \: r_{sw} \right) \:
      {\displaystyle
      \left( \frac{L_{vap}(T) \: q_{sw}}{R_{d}\:T_v}\right)}
          }
          { 1  \:  + \:
      \left( 1 + \eta \: r_{sw} \right) \:
      {\displaystyle
      \left( \frac{L_{vap}^2(T) \: q_{sw}}{c_p\:R_{v}\:T^2}\right)}
          }
          \: \right]
 \: .  \label{def_Gamma_sw_MG13}
\end{align}
It differs from the ``classical'' ones advocated by Durran and Klemp (1982) or Emmanuel (1994), which both contain an additional term in the denominator of the bracketered term, and do not return to $g/c_p$ when eliminating the aspects linked to condensation.

Probably because of the very general character of $\theta_s$, our saturated result on the contrary allows identifying as sole specific multipliers, the non-saturated adiabatic lapse rate, the ``full parcel kappa'' $R/c_p$ and the Clausius-Clapeyron factor. 
The results thus sound logical and especially consistent, since they take into account in a fully logical way the dependence of $L_{vap}$, $c_p$ and $R$ with the temperature and composition of the moist air.

 \section{New formulation for $\Gamma_{sw}$.} 
\label{section_new_Gamma_sw}

The saturated formulation (\ref{def_Gamma_sw_MG13}) can be rewritten as 
\begin{align}
  \Gamma_{sw} &  \: = \:
     \frac{g}{c_p}
          \: \left[ \: 
     \frac{ 1 \: + \:
             {\displaystyle
             \left( \frac{L_{vap}(T) \: r_{sw}}{R_{d}\:T}\right)}
          }
          { 1 \: + \: 
            {\displaystyle
            \left( \frac{R}{c_p} \right) \: 
            \left( \frac{L_{vap}(T)}{R_{v}\:T} \right) \: 
            \left( \frac{L_{vap}(T) \: r_{sw}}{R_{d}\:T} \right)}
          }
          \: \right]
 \: .  \label{def_Gamma_sw_new}
\end{align}

If trying to get away from our extreme cases (non-saturated and saturated), one notices that, owing to their simplicity, the transition between the non-saturated and saturated formulations (\ref{def_Gamma_ns}) and (\ref{def_Gamma_sw_new}) is equivalent to just replacing the constant ``1'' by $(R/c_p) [ L_{vap} / (R_v \: T) ]$. 

However, the second value may also be reorganised in the shape $[ L_{vap} / (c_p \:T) ] / (R_v / R)$. The latter expression is nothing else (Marquet and Geleyn, 2013, Appendix F) than the ratio of the impacts of water vertical transport on buoyancy, between saturated conditions (when only latent heat release acts) and non-saturated conditions (when only density-linked expansion acts).

Hence, defining by $C$ a weighting factor (which may, in a certain sense, be considered as the proportion of an air parcel being in saturated conditions), it is natural to express a {\it generalised shape for the vertical adiabatic lapse rate}, now under non-homogeneous conditions:
\begin{align}
F(C) & \; = \:
1 + C\: 
\left[\:
  \frac{L_{vap}(T)}{c_p\:T}\;
  \frac{R}{R_v}
  \; - \: 1 \:
\right] \: ,
\label{def_FC} \\
D_C & \; = \:
  \frac{L_{vap}(T)\:r_{sw}}{R_d\:T}
\: ,
\label{def_DC} \\
r_{sw} & \; = \: 
   \frac{\varepsilon \; e_{sw}(T)}
        {p \: - \:  e_{sw}(T)}
\: ,
\label{def_rsw}
\end{align}
leading to the compact result
\begin{equation}
\boxed{\;\;
\Gamma(C) \; = \;
     \frac{g}{c_p} \;
   \frac{1 + D_C}
        {1 + F(C) \: D_C}
 \;\; }
 \: .
\label{def_Gamma_C}
\end{equation}

Two remarks must be made.
\begin{itemize}[labelsep=5mm,listparindent=0mm,parsep=0cm,itemsep=1mm,topsep=0cm,rightmargin=2mm]
\item  Alike in the above-mentioned earlier publications, our definition of the saturation point corresponds to reversible conditions (i.e. at constant $q_t$) and not to the irreversible ones of ``permanent exact saturation''. In the second case it is $q_{sw}$ which would depend only on pressure and temperature, in the first case this happens for $r_{sw}$.
\item  If $D_C$ had been written with $r_v$ replacing $r_{sw}$, $\Gamma(C)$ would still have been compatible with its two extreme boundary conditions. 
But it is precisely in order to get the more logical situation of a term independent of the air parcel's composition multiplying $\Gamma(C)$ that we chose the above $D_C$  formulation for $\Gamma(C)$, expressed in terms of $r_{sw}$.
\end{itemize}

Concerning the second remark, one may even make $D_C$ more compact with the help of the Clausius-Clapeyron relationship:
\begin{align}
D_C & \; = \:
  \frac{L_{vap}(T)}{R_v\:T}   
  \left[ \:
  \frac{e_{sw}(T)}
       {p \: - \:  e_{sw}(T)}
  \: \right]
  \; = \;
  \frac{T}{p \: - \:  e_{sw}(T)} \;
  \frac{de_{sw}}{dT}
\: .
\label{def_DC_bis}
\end{align}

Most of this work was performed in the framework of the EU-ESF COST ES0905 action.

\vspace{5mm}
\noindent{\large\bf References}
\vspace{2mm}

\noindent{$\bullet$ Durran DR, Klemp JB.} {1982}.
{On the effects of moisture on the Brunt-V\"{a}is\"{a}l\"{a} Frequency.
{\it J. Atmos. Sci.}
{\bf 39} (10):
2152--2158.}

\noindent{$\bullet$ Emanuel KA.} {1994}.
{Atmospheric convection.}
Pp.1--580.
Oxford University Press: New York and Oxford.

\noindent{$\bullet$ Marquet P.} {2011 (M11)}.
{Definition of a moist entropic potential temperature. Application to FIRE-I data flights.
{\it Q. J. R. Meteorol. Soc.}
{\bf 137} (656):
768--791.
\url{http://arxiv.org/abs/1401.1097}
{\tt arXiv:1401.1097 [ao-ph]}}

\noindent{$\bullet$ Marquet P, Geleyn J-F.} {2013}.
{On a general definition of the squared Brunt-V\"{a}is\"{a}l\"{a} Frequency associated with the specific moist entropy potential temperature.
{\it Q. J. R. Meteorol. Soc.}
{\bf 139} (670):
85--100.
\url{http://arxiv.org/abs/1401.2379}
{\tt arXiv:1401.2379 [ao-ph]}}

\end{document}